\documentclass[pra
	       ,nofootinbib
	       ,floatfix
	       ,superscriptaddress
	       ,twocolumn
	       ]{revtex4-1}
	       
\usepackage{amsmath} 
\usepackage{graphicx} 
\usepackage{color} 
\usepackage[utf8]{inputenc}
\usepackage{siunitx}
\usepackage[hidelinks]{hyperref}

\newcommand{\fig}[1]{Fig.~\ref{fig:#1}}
\newcommand{\Fig}[1]{Figure~\ref{fig:#1}}
\newcommand{\refsec}[1]{Sect.~\ref{sec:#1}}

\newcommand{\um}[1]{\SI{#1}{\micro\meter}}

\newcommand{\ms}[1]{\SI{#1}{\milli\second}}

\providecommand{\abs}[1]{\ensuremath{\lvert#1\rvert}}

\begin{document}

\title{Scattering correcting wavefront shaping for three-photon microscopy}

\author{Bernhard~Rauer}
\email{bernhard.rauer@gmail.com}
\affiliation{Laboratoire Kastler Brossel, ENS-Université PSL, CNRS, Sorbonne Université, Collège de France, 24 rue Lhomond, 75005 Paris, France}

\author{Hilton B. de Aguiar}
\affiliation{Laboratoire Kastler Brossel, ENS-Université PSL, CNRS, Sorbonne Université, Collège de France, 24 rue Lhomond, 75005 Paris, France}

\author{Laurent Bourdieu}
\affiliation{Institut de Biologie de l’ENS (IBENS), École Normale Supérieure, CNRS, INSERM, Université PSL, Paris, France}

\author{Sylvain Gigan}
\affiliation{Laboratoire Kastler Brossel, ENS-Université PSL, CNRS, Sorbonne Université, Collège de France, 24 rue Lhomond, 75005 Paris, France}

\date{\today}

\begin{abstract}
Three-photon (3P) microscopy is getting traction due to its superior performance in deep tissues.
Yet, aberrations and light scattering still pose one of the main limitations in the attainable depth ranges for high-resolution imaging.
Here, we show scattering correcting wavefront shaping with a simple continuous optimization algorithm, guided by the integrated 3P fluorescence signal.
We demonstrate focusing and imaging behind scattering layers and investigate convergence trajectories for different sample geometries and feedback non-linearities.
Furthermore, we show imaging through a mouse skull and demonstrate a novel fast phase estimation scheme that substantially increases the speed at which the optimal correction can be found.
\end{abstract}

\maketitle

Light passing through biological tissue experiences an environment of heterogeneous refractive index, inevitably leading to aberrations and scattering~\cite{Ntziachristos2010}.
These effects limit optical microscopy to superficial tissue layers.
Adaptive optics (AO), originally developed for astronomy, allows to correct for common lower order aberrations, substantially improving imaging quality at depth~\cite{Hampson2021}.
In addition, point-scanning techniques that provide intrinsic optical sectioning, like confocal or two-photon (2P) microscopy, are likewise pushing the depth limit further into the tissue~\cite{Wilson2011,Helmchen2005}.

Still, light scattering poses one of the ultimate limitations for imaging deep inside living tissues.
It is, however, a coherent and deterministic process that can be controlled, even in regimes where no ballistic light is present~\cite{Vellekoop2007}.
To form a focus inside such strongly scattering environments non-invasively, one needs a metric providing feedback on the quality of the focus within the sample~\cite{Horstmeyer2015}.
In multi-photon fluorescence imaging, such a metric is conveniently given by the total fluorescence signal, due to the non-linear nature of the excitation process: a tight focus on a single target generates more signal than distributing the same power over multiple targets.
For 2P fluorescence (2PF), this feedback has been shown to facilitate focusing and imaging for a broad set of wavefront optimization algorithms~\cite{Katz2014,Papadopoulos2017,May2021,Blochet2021biorxiv}.

Recently, three-photon (3P) microscopy got adopted for biological imaging, primarily due to its superior performance at depth~\cite{Horton2013,Wang2018,Weisenburger2019}.
Both, the higher non-linearity of the excitation process and the shift towards longer excitation wavelength extends the achieved depth ranges.
Nevertheless, AO still proves highly useful to improve the 3PF signal when imaging at depth~\cite{Rodriguez2021,Streich2021,Sinefeld2022,Qin2022}.
This is because the higher non-linearity also leads to a higher susceptibility to signal loss due to aberrations.
Going beyond the correction of low-order aberrations is therefore hard, due the vanishing initial signal obtained from a strongly speckled point-spread function (PSF).
Even though, for particular optimization schemes, focusing and imaging has been demonstrated in such a context~\cite{Berlage2021}, a general understanding of the convergence criteria and optimal strategies for wavefront shaping with 3PF feedback are still missing.

Here, we demonstrate blind focusing behind a scattering layer through a simple continuous wavefront optimization scheme~\cite{Vellekoop2008a}, informed only by the total 3PF signal.
Scanning the obtained focus within the range of the memory effect~\cite{Feng1988}, we show signal improvements of up to two orders of magnitude and diffraction-limited imaging in situations where the unshaped PSF only resolves large-scale features.
Our approach reliably achieves focusing, also in volumetrically labeled three-dimensional (3D) samples.
For volumetrically labeled structures, it was recently argued that the enhanced non-linearity of 3P excitation is a prerequisite for convergence and that 2PF feedback could not achieve convergence to a focus in 3D samples \cite{Katz2014,Berlage2021}.
We therefore compare the convergence of continuous wavefront optimizations guided by 3PF and 2PF feedback for different sample geometries and show that 2PF feedback does indeed suffice to focus in volumetric targets.
We further demonstrate focusing and imaging through a thinned mouse skull bone, a scattering environment particularly relevant for neuroscience applications.
Finally, we implement a novel two-point phase estimation procedure, achieving a more than twofold improvement in speed for continuous  wavefront optimization schemes, highlighting the practicality of this simple approach.


\begin{figure*}[t]
    \includegraphics[width=1.0\textwidth]{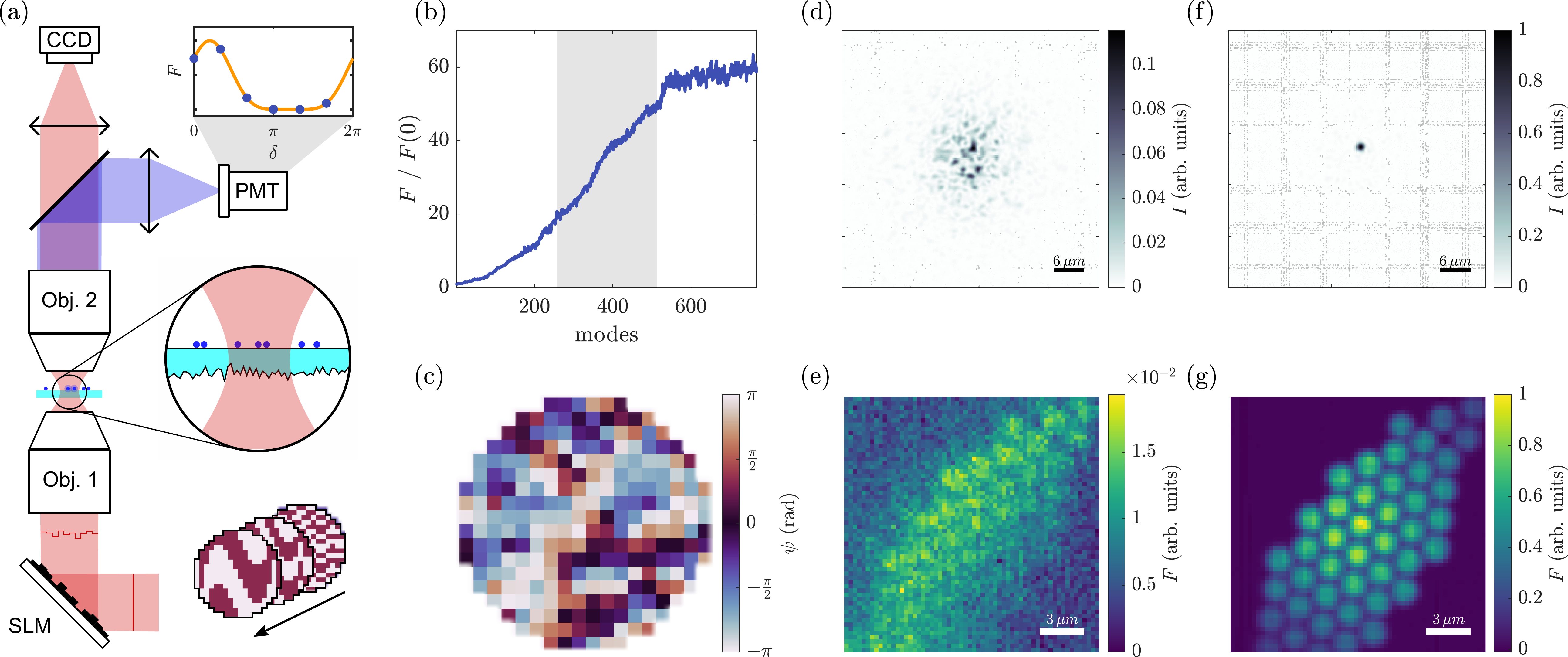}
    \caption{Continuous wavefront optimization guided by 3PF feedback. 
    (a) Scheme of the microscope setup. In the top right, the fluorescence signal recorded on the PMT when phase stepping a single mode is shown. Examples of circular Hadamard patterns are depicted in the bottom right. The central zoom-in shows the bead sample above the 2D scattering surface.
    (b) Evolution of the fluorescence signal enhancement during optimization. The gray and white backgrounds indicate a full set of 256 modes. 
    (c) Final correction pattern applied to the SLM.
    (d) PSF of the unshaped excitation beam, measured in transmission.
    (e) Image of the sample obtained from scanning the unshaped PSF shown in (d) over the sample.
    (f) PSF after wavefront shaping.
    (g) The corresponding scanned image.
    }
    \label{fig:fig1}
\end{figure*}

We use a custom-built microscope with a liquid crystal based spatial light modulator (SLM) conjugated to the back focal plane of the illumination objective (\fig{fig1}(a), for details see SM Sect. I).
A second objective aligned in transmission collects the emitted fluorescence which is measured on a photo-multiplier tube (PMT).
A CCD camera in the transmission path further enables monitoring of the excitation light pattern at the sample plane during the wavefront shaping procedure.
As samples, we use fluorescent beads which are \SI{2}{\micro\meter} in diameter.
The beads are drop-cast on a glass coverslip whose bottom surface, facing the illumination objective, was sand-blasted to introduce a single strongly scattering layer (see SM Sect. II).
This ensures a large speckle envelope and the absence of any ballistic light.

To optimize the incident wavefront we use a continuous optimization algorithm~\cite{Vellekoop2008a}.
In each optimization step, half of the active pixels on the SLM are phase-stepped between 0 and 2$\pi$ while the other half is kept constant and acts as a reference.
From the observed 3PF signal, we extract the phase at which the modulated pixels are optimal and add it to the corresponding pixels of the current optimal pattern.
The resulting pattern forms the initial state of the next optimization step and is thereby continuously improved.
The mode patterns, determining the active pixels in each optimization step, are chosen to be circular Hadamard patterns (\fig{fig1}(a) bottom right, for details see SM Sect. III).
While more involved procedures have been used for focusing through wavefront shaping with non-linear feedback~\cite{Papadopoulos2017,May2021}, this algorithm represents one of the most simple ones.

A typical evolution of the 3PF signal under optimization is shown in \fig{fig1}(b).
Here, the active area on the SLM was decomposed into 256 modes which were optimized three times, consecutively.
We see that already after the second round the fluorescence signal almost plateaued to its maximal value, a number of measurements comparable to other approaches~\cite{Berlage2021}.   
The final SLM pattern (\fig{fig1}(c)) reaches a 60-fold signal enhancement, realizing a tight focus of the excitation light at the target plane (\fig{fig1}(f)).
Note that, while the maximal intensity of the excitation light increases by about a factor of 10, the fluorescence signal enhancement stays way below the cube of this value.
This is because the target sample is extended such that the initial speckled PSF excites many beads at once and the redistribution of power into a single focus reduces the signal collected from neighboring beads.

Applying additional linear phase ramps to the SLM allows us to scan the beam over the sample without dedicated scanning mirrors, albeit at a lower speed.
Within the memory effect range of the scattering geometry, the optimized correction pattern stays valid and we can form an image of the hidden sample.
Additionally, laterally translating the correction pattern on the SLM proportionally to the phase gradients applied for each position extends the validity range of the correction further~\cite{Osnabrugge2017,Papadopoulos2020} (see SM Sect. IV).
The resulting image with the beads well resolved can be seen in \fig{fig1}(g).
In contrast, in an image formed by scanning the unshaped PSF over the sample we could only resolve global shapes (\fig{fig1}(d) and (e)).
This is different to many multi-photon AO implementations where often a substantial signal enhancement is achieved while improvements in resolution remain small.

Due to the cubic non-linearity of the feedback signal and the multitude of sources illuminated by the initial speckle pattern it is not a given that the optimization converges to a single focus.
Yet, we observe reliable focusing on similar time scales for many different speckles and sample geometries, even though the exact position of the focus cannot be predicted a priori.
The question how the convergence to a focus depends on the non-linearity of the feedback process and the geometry of the sample is a point of interest in recent studies on this topic.
It was argued that for volumetrically labeled samples 2PF feedback does not suffice to form a single focus~\cite{Katz2014,Berlage2021}.
The argument is based on the observation that under 2P excitation the signal collected from a Gaussian excitation beam inside a volumetric dye pool does not depend on the NA of the beam.
Only for a third order non-linearity this dependence appears.
Nevertheless, in the context of 2P AO, convergence was observed for aberrated beams in volumetric samples~\cite{Sinefeld2015,Tang2012}.
For a 2D geometry, it was also predicted that for certain optimization schemes there exists a maximal number of targets such that wavefront shaping under non-linear feedback reliably converges to a focus~\cite{Osnabrugge2019}.

\begin{figure}[t]
    \includegraphics[width=1.0\columnwidth]{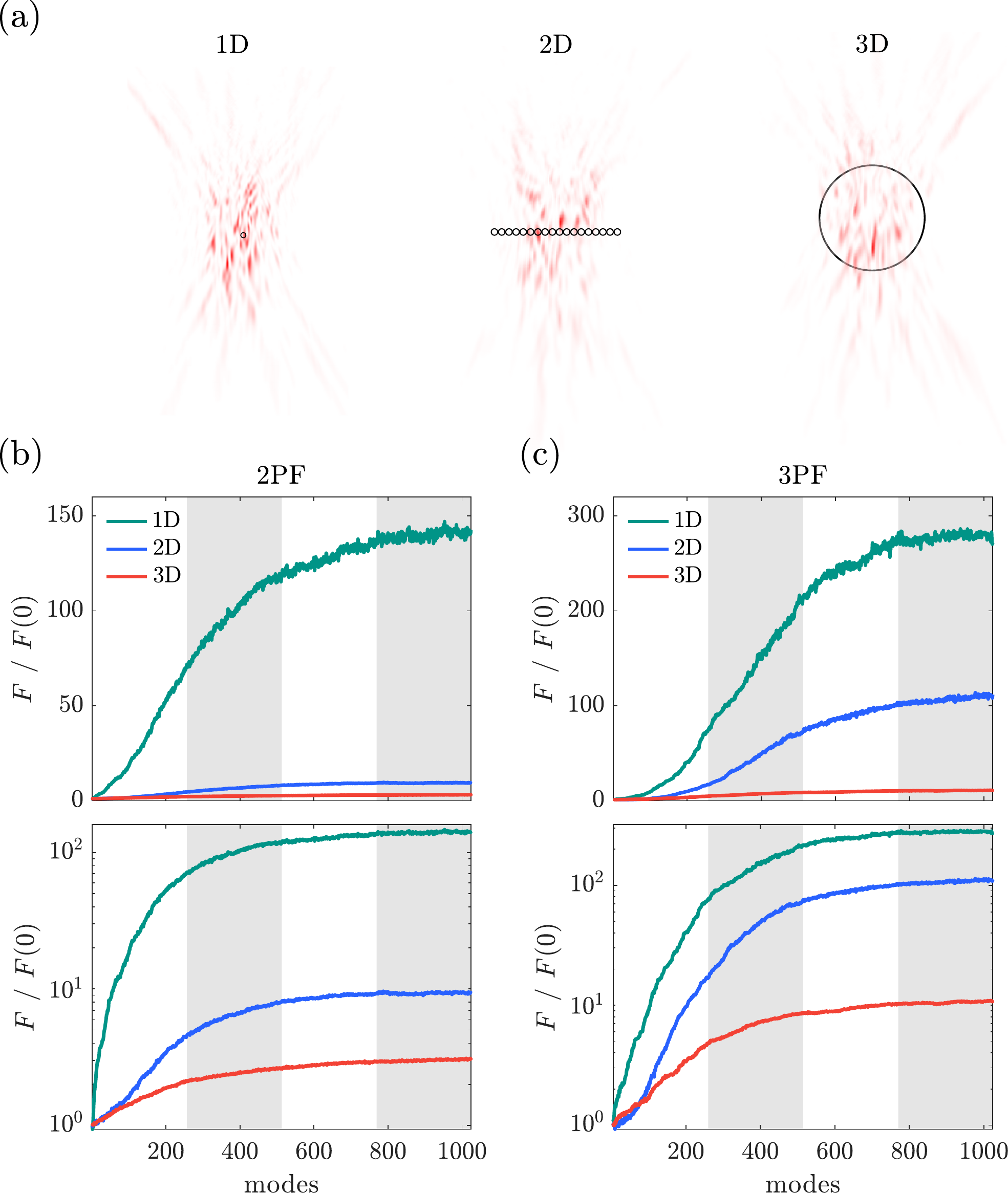}
    \caption{
	Convergence study.
	(a) Schematic of the sample geometries. 1D: an isolated bead \um{2} in diameter, 2D: a homogeneous layer of the same beads, 3D: a large bead \um{15} in diameter.
	(b) The upper panel shows a comparison of signal enhancement during wavefront shaping for different geometries under 2P feedback. The gray and white areas indicate single repetitions of the set of 256 modes used. The lower panel shows the same data in log scale.
	(c) The same for 3P feedback.
	All curves are averaged over at least four realizations of the scattering layer.
    }
    \label{fig:fig2}
\end{figure}

We therefore investigate the convergence properties of the wavefront shaping algorithm for different sample geometries under 2P and 3P excitation.
For that, we prepare samples with an effective 3D, 2D, and 1D distribution of fluorophores (\fig{fig2}(a)).
The 1D samples consist of a single bead while 2D samples are created from bead monolayers which allows for a simple preparation with a precise thickness.
For the 3D samples we use large beads with a diameter of \SI{15}{\micro\meter}. 
Being about the size of a neuron soma, such finite volumetric targets are much more relevant to biomedical applications than dye pools.
Also, when interested in geometric effects, they assure that convergence is not hindered by the total number of targets~\cite{Osnabrugge2019}.

Figure~\ref{fig:fig2}(b) and (c) show the signal enhancement under optimization for 2P and 3P excitation, respectively.
We observe that for 2P excitation, the difference in enhancements between 1D and extended targets is much larger than in the case of 3P excitation.
We attribute this to the geometric effects discussed above.
Interestingly, comparing the curves for the different geometries in log scale, we see no large differences in convergence time within one excitation regime.
In all cases, the final signal level is reached after about three optimization runs through the full mode basis.

\begin{figure}[t]
    \includegraphics[width=1.0\columnwidth]{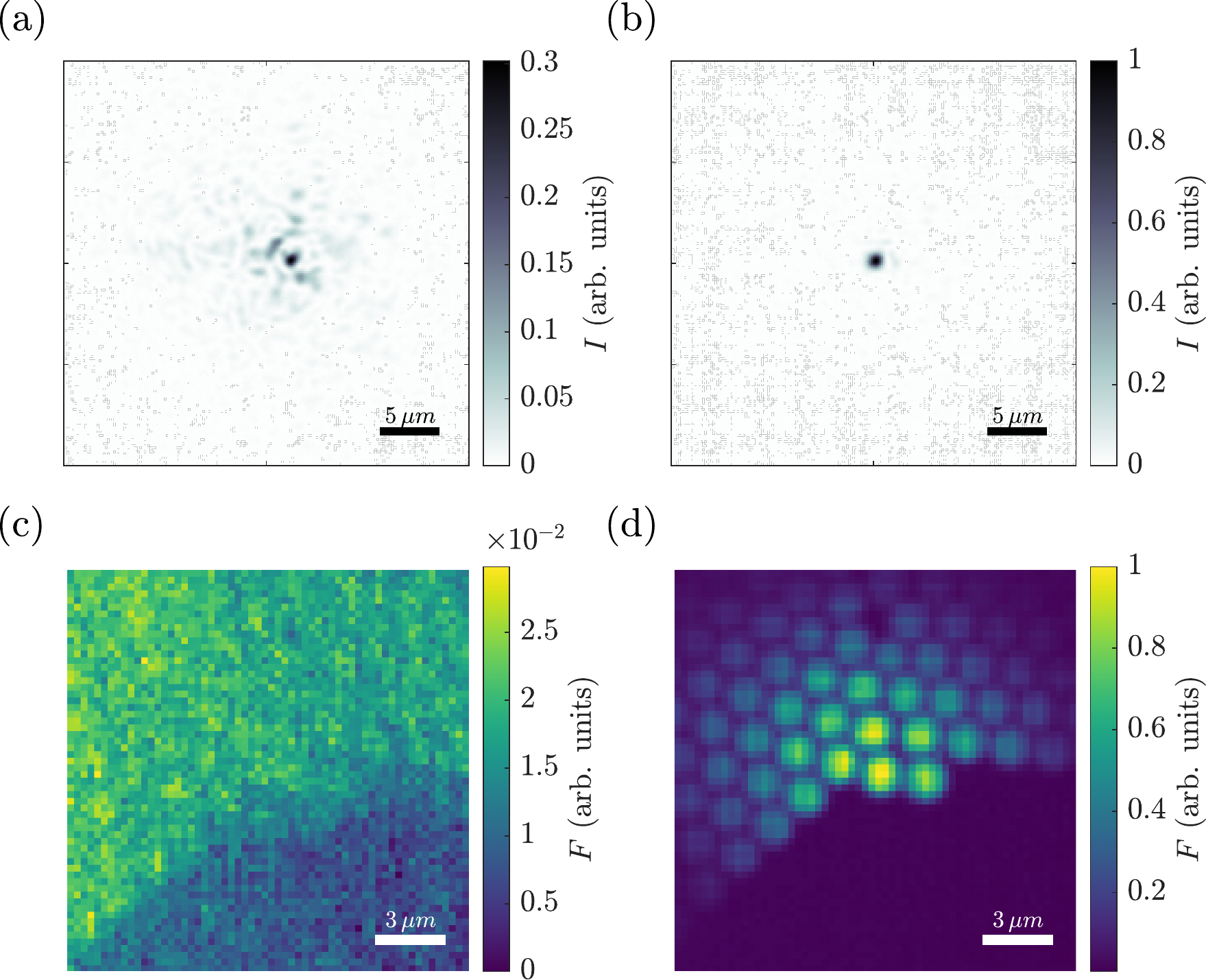}
    \caption{
    Focusing through a thinned mouse skull bone. 
    (a) Initial speckled PSF before wavefront shaping. 
    (b) Focus obtained after optimizing 256 modes, each two times. 
    (c) Image obtained from scanning the PSF shown in (a) over the bead sample.
    (d) Corresponding image formed by scanning the shaped PSF.
    }
    \label{fig:fig3}
\end{figure}

We further observe that 2P feedback does indeed lead to a single focus within the large bead volume.
Even though the signal enhancement is comparatively small, the algorithm converges reliably. 
This points to the conclusion that there is no qualitative difference for 2P feedback in volumetric samples.
Yet, as the increase per mode is small, for some practical situations, noise levels can render 2P feedback ineffective.
While it could be argued that the finite volume used here aids convergence, we also observe focusing in dye pool samples (see SM Sect. V).
Note that, even though the small 2P signal increase for volumetric samples renders wavefront shaping ineffective in weakly scattering regimes, in the case of a fully speckled PSF with no strong ballistic component the situation is different. 
There, correcting the wavefront enables imaging, even if the initial uncorrected signal is of approximately the same magnitude

To show the practical applicability of this wavefront shaping technique we use it to focus through a relevant biological obstacle: a mouse skull bone.
The skull was thinned to about \SI{200}{\micro\meter} and fluorescent beads were placed behind it.
Figure~\ref{fig:fig3}(a) and (b) show a PSF measured in transmission before and after shaping, demonstrating focusing after twice optimizing 256 modes.
The images obtained with these PSFs show a similar improvement as for the single scattering layer, albeit with a slightly smaller total enhancement and a more limited memory effect range (\fig{fig3}(c) and (d)). 
This is expected as the skull bone is 3D scattering obstacle, compared to the 2D ground glass layer.

\begin{figure}[t]
    \includegraphics[width=1.0\columnwidth]{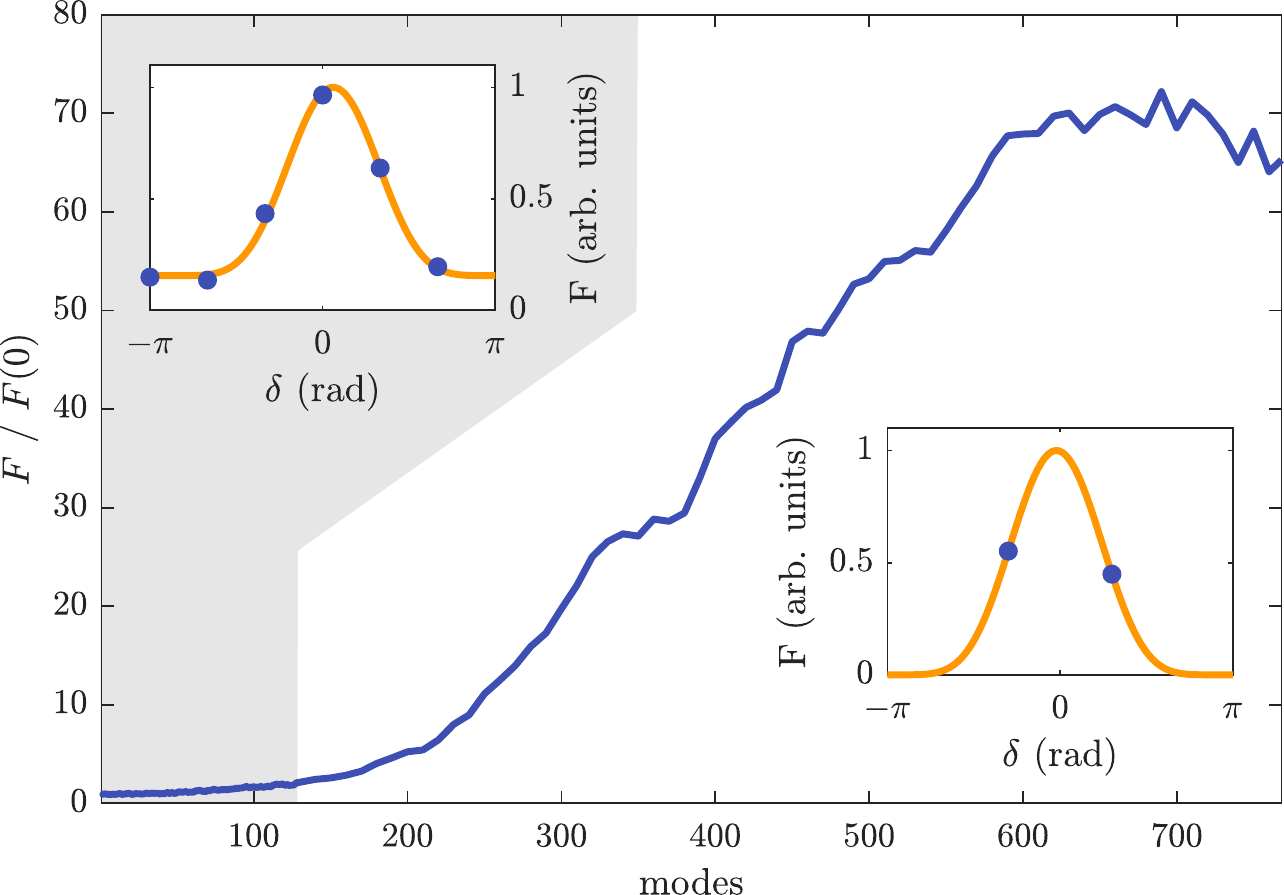}
    \caption{
 	Fast phase estimation.
 	Evolution of the fluorescence signal enhancement for an optimization using 6 phase steps for the initial 128 modes and 2 phase steps thereafter.
 	The two insets show examples of phase estimation (blue dots) for the 6-point (top left) and 2-point measurements (bottom right).
 	The orange lines show the expected non-linear fluorescence signal with the phase extracted from the corresponding methods.
    }
    \label{fig:fig4}
\end{figure}

The downside of a modulation scheme where in each iteration half the SLM pixels are phase stepped while the other half acts as a reference is the pronounced non-linearity of the feedback signal.
Even though it leads to the maximal signal contrast, this configuration requires more than 4 measurements per mode, especially for 3rd order non-linearities or above (see SM Sect. VI).
For the measurements presented so far we therefore used 6 measurement points to determine the optimal phase for each mode.
However, when monitoring the estimated phases during the optimization one realizes that they quickly tend to zero after the first few iterations.
This is because the applied wavefront starts to converge to the optimal one, rendering each further improvement incremental.
Knowing this we can devise an optimal measurement strategy that does not require a full 6 measurements for each iteration (see SM Sect. VII).
Choosing two measurement points that coincide with the maximal gradient of the non-linear fluorescence signal allows us to estimate the optimal phase using far less measurements.
This is reminiscent of the standard 3-point measurements scheme used in AO~\cite{Debarre2007,Sinefeld2015}, only that here we know the functional shape of the signal and can precisely determine the optimal stroke with just two data points.
\Fig{fig4} shows the application of this technique, demonstrating a speedup of more than a factor of two for the full optimization procedure.
Note that, at the point in the optimization where the fast 2-point phase estimation starts to be applied, the signal is only 3\% of its later maximal value.

Recently, several novel wavefront shaping strategies aimed at focusing inside scattering environments have been developed~\cite{Papadopoulos2017,May2021}.
Here, we demonstrate that for 3PF samples a simple continuous wavefront shaping procedure suffices to blindly focus behind a scattering layer, even in the absence of any initial ballistic component.
It reliably converges and can even be run at two measurements per mode, after some initial pre-optimization.
We show imaging in situations where initially only large structures can be resolved and further demonstrate its application with biologically relevant scattering media. 
Investigating the convergence in continuous optimization for different target geometries using both 2PF and 3PF feedback we further demonstrate the advantage a higher non-linearity gives in the signal increase obtained for extended sources.
We also show that, contrary to the intuition obtained from a Gaussian beam model, 2PF feedback does allow focusing inside homogeneous volumetric targets, albeit with low signal enhancements.
This highlights that wavefront optimization based on non-linear feedback is far from trivial and we hope the results presented here inspire further numerical and experimental studies.

\vspace{5mm}
\noindent{\bf Acknowledgments:} \\
This project was funding by the European Research Council under the grant agreement No. 724473 (SMARTIES), the European Union's Horizon 2020 research and innovation program under the FET-Open grant No. 863203 (Dynamic) and the program "Investissements d’Avenir" launched by the French Government and implemented by ANR with the references ANR–10–LABX–54 MEMOLIFE and ANR–10–IDEX–0001–02 PSL* Université Paris.
B.R. was supported by the European Union's Marie Sk\l{}odowska-Curie fellowship, grant agreement No. 888707 (DEEP3P).

The authors thank Walther Akemann for providing the mouse skull; Julien Guilbert and Gerwin Osnabrugge for useful discussion and and Fei Xia for the careful reading of the manuscript.

\bibliography{Papers-3PF,Papers-3PF_add_lib}

\clearpage
\onecolumngrid

\renewcommand{\thefigure}{S\arabic{figure}}
\setcounter{figure}{0}

\begin{center}
  \LARGE
  \textbf{Supplementary Materials} 
\end{center}

\section{Experimental setup}

\noindent The experimental setup is sketched in Fig. 1(a) of the main text.
The source used is an OPO (Coherent Mira OPO-X) pumped by a pulsed Ti-sapphire laser (Coherent Chameleon Ultra II) with \SI{140}{\femto\second} pulse duration and repetition rate of \SI{80}{\mega\hertz}.
To modulate the wavefront, we use a liquid crystal phase-only SLM (Meadowlarks HSP512L) which is imaged on the back focal plane of a water immersion objective (Zeiss W Plan-Apochromat 40x/1.0 DIC M27). 
After illuminating the sample, a second microscope objective (Zeiss EC Plan-NEOFLUAR 40x/1.3) images the excitation light pattern onto a CCD camera (Basler acA1300-30um) and collects the fluorescence light which is diverted towards a PMT (Hamamatsu H7422P-40) by a dichrocic mirror (Semrock Di03-R785-t1-25x36).
Additional filters in front of the PMT (Thorlabs FESH0600, FESH0700) assure that only fluorescent light is detected.

For 3P excitation, the OPO is tuned to \SI{1050}{\nano\meter}.
As targets, we use fluorescent beads with a single photon absorption peak in the blue that are \SI{2.1}{\micro\meter} in diameter (Thermo-Fisher Fluoro-max B0200 - 365, 388, \SI{412}{\nano\meter} / 445, 445, \SI{473}{\nano\meter}).
The large beads used as 3D targets for the measurements presented in Fig. 2 of the main text are \SI{15}{\micro\meter} in diameter and have a similar single photon absorption peak (Invitrogen F8837 - \SI{365}{\nano\meter} / \SI{415}{\nano\meter}). 
For the 2P excitation measurements presented in Fig. 2 of the main text, we use the same beads excited directly by the light from the pump laser at \SI{800}{\nano\meter}.

Typically, for 3P excitation, low repetition rate lasers are used, as for a repetition rate of \SI{80}{\mega\hertz}, the 3P the absorption is quite inefficient.
We therefore use initial laser powers of about \SI{180}{\milli\watt} at the sample plane and detect the fluorescence in transmission with a high NA oil immersion objective.
Note that, the choice against detecting fluorescence in an epi-geometry is solely based on the need to increase signal levels and is no requirement of the optimization strategy nor the imaging technique used.
In addition, we underfill the illumination objective back aperture to prevent clipping and the associated power loss.
The Gaussian beam entering the aperture has a $1/e^2$ width of \SI{6.3}{\milli\meter} for the 3P excitation at \SI{1050}{\nano\meter} and \SI{8}{\milli\meter} for the 2P excitation at \SI{800}{\nano\meter}.
The back aperture of the illumination objective is about \SI{9}{\milli\meter} wide.
On the SLM, we modulate a circular region (see Sect. \ref{sec:circ_had}) that spans about 80\% of the $1/e^2$ beam width, for both 3P and 2P excitation.

To exclude the correction of aberrations inherent to the optical setup, we measure the optimal system correction pattern and use it as the initial point for the wavefront optimization process.
For that, we use samples with a single fluorescent bead of diameter \SI{0.5}{\micro\meter} (Invitrogen F8812 and F8813).
These bead samples are prepared in the same way as the samples used for the measurements presented in the main text (see \refsec{samples}) but without a scattering layer.
Starting from a flat SLM pattern and optimizing the total non-linear fluorescent signal when focusing on one of these beads we obtain the system correction pattern.
For higher signal strength, we use a 2P excitation for both the \SI{1050}{\nano\meter} and the \SI{800}{\nano\meter} light.
The final pattern serves as the initial wavefront for the optimization procedures presented in the main text.
Note that, the signal increase due to the system correction is maximally 25\% and therefore negligible compared to the one achieved in scattering correction.

In our setup, the speed of the wavefront optimization is limited by the integration time applied to the PMT signal, which is set to \ms{40}.
Again, this is necessary due to the low signal levels.
With the \ms{10} it takes to load a new SLM pattern and some processing time and additional delays that assure smooth operation, this results in \ms{70} for a single data point and \ms{420} to perform a 6-point phase stepping measurement.
In the case of a 2-point phase estimation, as presented in Fig. 4 of the main text, the time to measure one mode reduces to \ms{140}.
Note that, using a source more adapted to 3P excitation would reduce this time substantially.

The signal increase during wavefront shaping can be large, up to 280-fold in the case of a single bead under 3P excitation, as demonstrated in Fig. 2(c) of the main text.
To prevent bleaching and saturation of the PMT we therefore reduce the laser power during the optimization.
This is automated with a continuously variable neutral density filter (Thorlabs NDC-50C-4) rotated by a computer controlled stepper motor.
The power is reduced as soon as the threshold of a 7 to 10-fold signal increase is reached. 

\section{Sample preparation}
\label{sec:samples}

\noindent A scattering surface is introduced to a \#{}1.5 cover slip by blasting its bottom surface with 220 grit sand.
To reduce the scattering, the resulting rough surface is covered with Norland NOA65 glue which is immediately wiped off with a lens tissue.
After that, the glue remaining on the surface is cured under UV light. 
The beads are drop-casted on the top surface of the coverslip, about \SI{170}{\micro\meter} above the scattering layer. 
They are covered by the same NOA65 glue to create a near-index matched environment.
Finally, a second \#{}1.5 cover slip is placed on top of the beads and the glue is cured.
This second curing process is kept short not to bleach the beads.
Near-index matching the polystyrene beads ensures that the excitation light recorded in transmission is not distorted by the beads.

For the measurement presented in Fig. 3 of the main text, a piece of the parietal bone of a male 7.5 weeks old C57BL/6 mouse was used.
The bone was thinned down to $\sim$\SI{200}{\micro\meter}, immersed in phosphate-buffered saline and sandwiched between two \#{}1.5 cover slips separated by a \SI{240}{\micro\meter} thick spacer.
The beads are drop-cast on the top surface of the upper coverslip and covered by UV-cured glue in the same procedure as described above.

Experimental procedures were conducted in accordance with the institutional guidelines and in compliance with French and European laws and policies.
They were approved by the ‘Charles Darwin’ local institutional ethical committee registered at the French National Committee of Ethical Reflection on Animal Experimentation under the number 05 (authorization number: APAFIS 26667)

\section{Circular Hadamard modes}
\label{sec:circ_had}

\noindent The rows of the Hadamard matrix, reshaped into 2D patterns, are a common choice of basis for wavefront shaping applications beyond Zernike polynomials.
They are orthogonal and each mode divides the pixels in the modulated area into two equally large sets.
The latter property assures a maximal interference signal when modulating one of the sets against the other (see Sect. \ref{sec:phase_est}) while the former provides a faster convergence in an iterative optimization scheme compared to, e.g., a random basis.

\begin{figure}[t]
	\centering
    \includegraphics[width=0.7\textwidth]{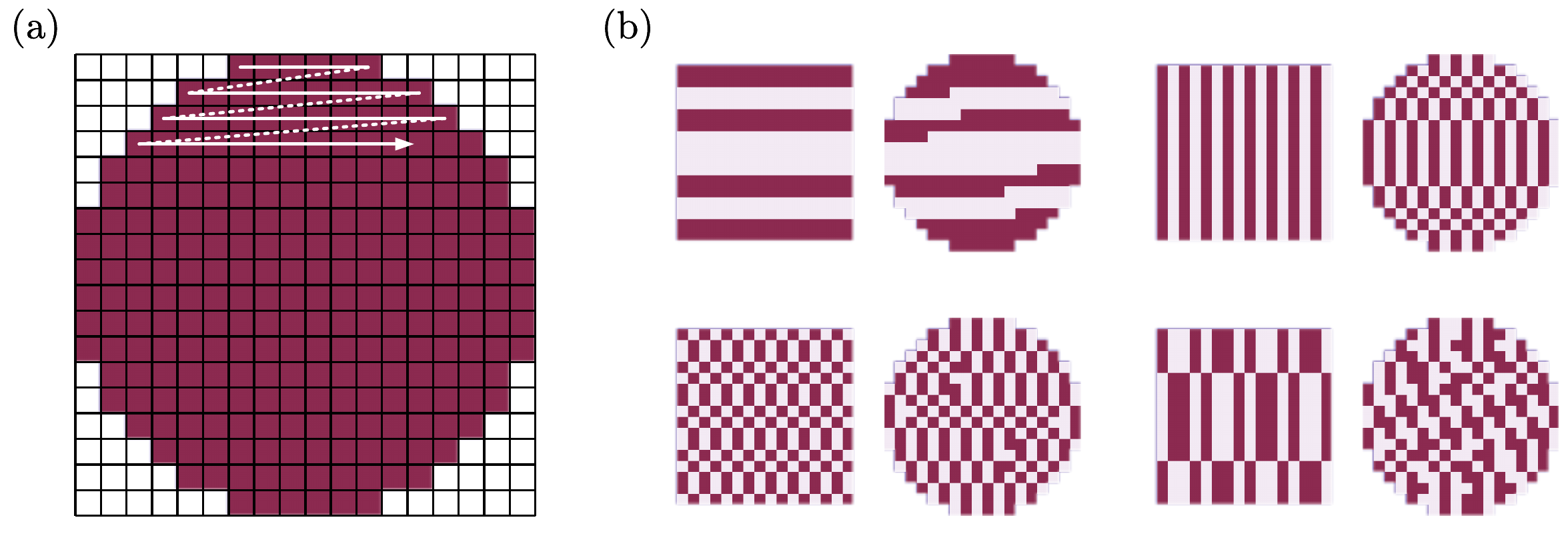} 
    \caption{
    Circular Hadamard modes. 
    (a) Circular pixel layout for $N_\mathrm{mode} = 256$. The rows of the Hadamard matrix are arranged following the white arrow.
    (b) Four examples of square Hadamard modes and their circular counterparts (pairs left and right, respectiveley).
    }
    \label{fig:figS_circHad}
\end{figure}

In microscopy, however, we often want to control the light field at the backfocal plane of an objective or at some other circular symmetric optical component.
Here, the square shaped Hadamard modes are sometimes unpractical as they can be clipped by back apertures or lead to an asymmetric modulation of the light.
The square shape though is arbitrarily chosen and not inherent to the originally 1D rows of the Hadamard matrix.
We can arrange the rows in any kind of 2D pattern while still preserving the patterns' orthogonality and equal-division of pixels.
A circular arrangement for example, as shown in \fig{figS_circHad}(a), can be optimal to when working with circular optics and apertures.
The resulting mode patterns are slightly wider than their square equivalents, as shown for two examples in \fig{figS_circHad}(b). 

Note that, while for $N_\mathrm{mode} = 256$ a perfect circle can be arranged, for other basis set sizes a few pixels have to be added or removed asymmetrically to assure a number of pixels equal to a power of two, as required by the Hadamard matrix.
These few pixels, however, generally do not affect the circular shape much.

\section{Shift scanning}

\begin{figure}[t]
	\centering
    \includegraphics[width=0.75\textwidth]{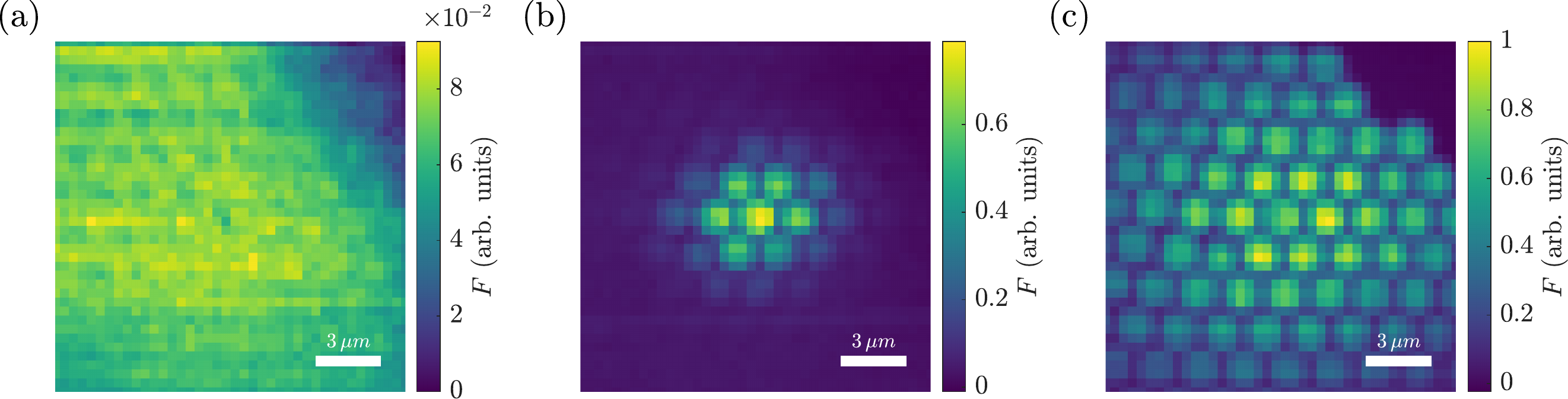} 
    \caption{
    Shift-enhanced memory effect range.
    (a) 2PF image obtained from scanning a speckled PSF over a bead target.
    (b) Image obtained after wavefront shaping, applying phase gradients to the SLM to move the focus over the sample.
    (c) Same image when, in addition to the phase gradients, the optimal correction mask is shifted in the direction of the gradient. 
    }
    \label{fig:figS_shift}
\end{figure}

\noindent To scan the optimized focus over the sample, forming an image, we apply phase ramps to the SLM.
This is a very slow imaging technique which could be easily sped up by introducing galvanometric mirrors to the beam path after the SLM.
However, performing the scan with the SLM provides the advantage of being able to correct the applied shaping pattern while scanning.
For example, one can partially correct the fact that while scanning, the scattering regions the beam traverses shift with respect to beam, misaligning the correction pattern and the scatterers~\cite{Osnabrugge2017,Papadopoulos2020}.
Simply shifting the correction pattern on the SLM as a function of the applied phase gradient leads to a substantially enlarged memory effect range, as shown in \fig{figS_shift}.
Here, the optimal amount of shift is found iteratively, by tuning the parameter while observing the range at which the wavefront correction is valid.
For the images presented in the main text a shift per phase gradient of \SI{0.032}{\milli\meter^2\per\radian} was used.
At the edge of the image this corresponds to a 19 pixel shift, which is about 10\% of the total pattern length.
With the thin 2D scattering layers, we observed an extension of the memory effect range of excitation intensity by a factor of 2.7.
For volumetric scatterers, like the mouse skull bone of Fig. 3 of the the main text, the gain in imaging range is smaller.

\section{Dye pool measurement}

\begin{figure}[t]
	\centering
    \includegraphics[width=0.75\textwidth]{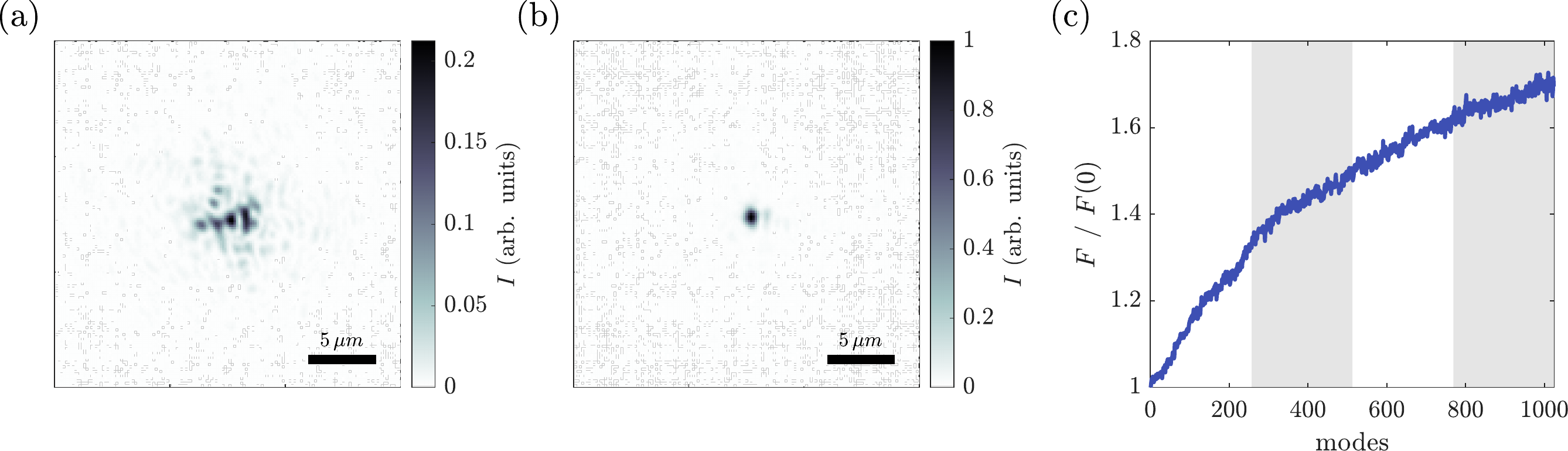} 
    \caption{
    Focusing guided by 2PF feedback inside a dye pool. 
    (a) PSF of the unshaped excitation beam, measured in transmission.
    (b) PSF after wavefront shaping.
    (c) Evolution of the fluorescence signal enhancement during optimization. The gray and white backgrounds indicate a full set of 256 modes. The enhancement is averaged over three scattering realizations.
    }
    \label{fig:figS_dyePool}
\end{figure}

\noindent To assure that the convergence to a focus with 2PF feedback observed in the measurements presented in Fig. 2 of the main text is not due to the finite size of the fluorescent target we also performed the same measurement on a dye pool sample with a thickness of \um{120}.
The results are presented in \fig{figS_dyePool}.
The initial speckle converges towards a clear focus although the increase in total fluorescence is only 70\%.
Also, the convergence is slower than observed in the measurements of the main text, even after sampling through all modes four times the wavefront does not seem to have fully converged.

Note that, for these measurements we needed to reduce the size of the initial speckle envelope to observe convergence.
We believe this is related to the limits encountered when exciting a large number of targets, discussed in \cite{Osnabrugge2019}, although 3D targets are not discussed there.

To reduce the scattering and thereby shrink the envelope we use a glue at a lower refractive index (Norland NOA85) and fully cover the sand blasted cover glass surface without wiping it off.
We then add an additional coverslip on the bottom and cure the glue.
This reduces the refractive index difference at the rough interface, limiting the scattering to smaller angles.

\section{Phase estimation}
\label{sec:phase_est}

\noindent Tuning the phase of a given mode during the optimization procedure leads to a total $n$-photon fluorescent signal
\begin{equation}
\label{tot_fluor}
F(\delta) \propto \sum_i \sigma_{i,n} \big[ \abs{E^{i}_\mathrm{ref}}^2 + \abs{E^{i}_\mathrm{mode}}^2 + 2 \abs{E^{i}_\mathrm{ref}} \abs{E^{i}_\mathrm{mode}} \mathrm{cos}(\varphi_{i} + \delta) \big]^n.
\end{equation}
Here, $E^{i}_\mathrm{mode}$ is the field of the tuned mode at the point of source $i$ while $E^{i}_\mathrm{ref}$ is the reference field at that same point.
The phase between these two fields is given by $\varphi_{i}$ and $\delta$ denotes the applied phase stepping.
The signal is detected on a bucket detector such that the sum runs over all sources $i$ that are illuminated, with $\sigma_{i,n}$ being the $n$-photon cross section of each source.

Typically, in continuous wavefront optimization schemes, the optimal phase of a single mode is determined by stepping $\delta$ from 0 and $2\pi$ at equal spacing.
Assuming an offset cosine modulated signal and enough measurement points, the optimal phase is then given by 
\begin{equation}
\label{phi_opt}
\varphi_\mathrm{opt} = \mathrm{angle} \Big[ \sum_\delta F(\delta) e^{i\delta} \Big].
\end{equation}
However, the cosine shape of $F(\delta)$ is only assured for linear feedback mechanisms.
For $n>1$, the functional form of the total fluorescent signal depends on the target and the amplitude balance between reference and mode fields.
Expanding the $n$-th power in eq.~\ref{tot_fluor}, we can see that only if one of the two fields is much larger than the other a term linear in $\mathrm{cos}(\varphi_{i} + \delta)$ dominates the $\delta$ dependent terms.
Therefore, it is a viable strategy to choose $\abs{E^{i}_\mathrm{ref}} \ll \abs{E^{i}_\mathrm{mode}}$ or vice versa in order to assure a cosine shaped modulation signal.
The downside is a small modulation amplitude compared to the background fluorescence.

\begin{figure}[t]
	\centering
    \includegraphics[width=0.9\textwidth]{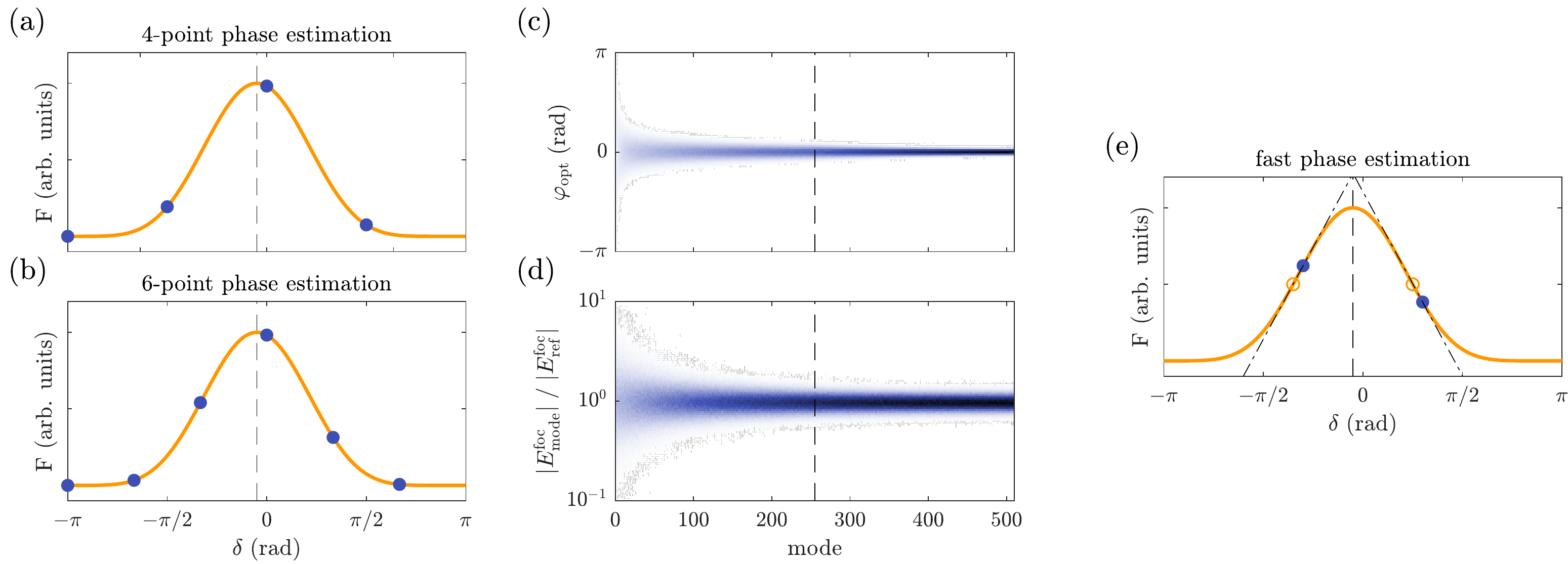}
    \caption{Phase estimation for continuous wavefront shaping. 
    (a) Phase estimation with 4 equally spaced measurement points (blue dots) for a maximally non-linear feedback signal originating from a single 3PF source (orange line).
    (b) The same with 6 equally spaced measurement points.
    (c) Simulated distribution of $\varphi_\mathrm{opt}$ for a continuous wavefront optimization. The scattering is modeled by a complex Gaussian i.i.d. transmission matrix with 256 input modes. This set of modes is optimized two times with the dashed vertical line indicating the two sets. The 3PF target is extended and covers 27\% of the output pixels.
    (d) Amplitude ratio of the currently optimized mode and the reference field at the focal point for the same simulation.
    (e) Phase estimation with 2 measurement points at $\tilde{\delta}_\pm$. The dash-dotted lines with slopes $\pm\;\alpha/2$ indicate the first order phase estimation. The orange circles show the large signal changes even for $\varphi_\mathrm{opt} \simeq 0$. For (a), (b) and (e) $\varphi_\mathrm{opt} = 0.05\;\pi$.
    }
    \label{fig:fig_fpe}
\end{figure}

For $\abs{E^{i}_\mathrm{ref}} = \abs{E^{i}_\mathrm{mode}}$ the modulation amplitude is the largest, but for a single target $F(\delta) \propto \big[1 + \mathrm{cos} ( \varphi_\mathrm{opt} + \delta) \big]^n$ and for multiple fluorescent sources the superposition of many non-linear modulations leads to strongly target dependent functional forms. 
Yet, eq.~\ref{phi_opt} still provides a good estimate of $\varphi_\mathrm{opt}$ in this case and a continuous wavefront optimization based on it converges quickly.
A critical point, however, is the number of phase steps acquired for each mode.
While typically a 4-point phase stepping is used, for non-linear fluorescence signals this can be insufficient, as shown in \fig{fig_fpe}(a) for a 3PF signal from a single target.
For certain optimal phases, the measured points are sub-optimally placed in regions of flat or slowly varying signals leading to increased errors in the estimation of $\varphi_\mathrm{opt}$.
This can be counteracted by measuring $F(\delta)$ at more points, as shown in \fig{fig_fpe}(b), with the downside of a reduced optimization speed.

For a continuous wavefront optimization procedure using Hadamard modes where in each step half of the SLM pixels are optimized and the other half acts as the reference, the two interfering fields are on average balanced leading to non-cosine modulations of the total fluorescence signal. 
Therefore, for the data presented in Fig. 1, 2 and 3 of the main text, a 6-point optimization scheme is used.

Another interesting point to note is that for a continuous optimization strategy where always half of the pixels are modulated, the reference field changes in each iteration.
This leads to a slower conversion, as pixels are constantly optimized relative to a different reference.
Even in case of a linear feedback signal, such an optimization does not fully converge after sampling through all modes once.
The requirement to optimize each mode 2-3 times, observed in Fig. 1(b), Fig. 2 and Fig. 4 of the main text, is therefore not solely a result of the non-linear feedback signal.

\section{2-point phase estimation}
 
\noindent For continuous wavefront optimization schemes, the current wavefront gets closer and closer to one that focuses on a single target as the optimization continues.
Also, the partition in the modulated part of the wavefront and the reference part changes in each iteration.
This means, as the optimization continuous, both the current mode and the reference field tend to be already close to their optimal orientation even before the phase stepping procedure starts.
We can observe this when looking at the distribution of the extracted optimal phases or when comparing the amplitudes of the current mode and the reference at the brightest spot during the optimization (see \fig{fig_fpe}(c) and \fig{fig_fpe}(d), respectively, for simulated distributions).
Together with the observation that even for complicated and large targets the procedure quickly converges to a single focus which is then slowly optimization, this leads to predictable fluorescence signal after a few iterations.
The balance between $\abs{E_\mathrm{ref}}$ and $\abs{E_\mathrm{mode}}$ forces $F(\delta) \propto \big[1 + \mathrm{cos} ( \varphi_\mathrm{opt} + \delta) \big]^n$, as discussed above, and $\varphi_\mathrm{opt}$ is always close to zero.

This knowledge can be used to selectively measure only at those phase points that provide the most information and thereby speed up the optimization.
From \fig{fig_fpe}(b) for example, we see that the two points in the flanks of the peak will be most crucial in determining $\varphi_\mathrm{opt}$.
Ideally, we want to measure at the maximal signal gradient, which occurs at 
\begin{equation}
\bigg[\frac{1 + \mathrm{cos} (\tilde{\delta})}{2} \bigg]^n = \frac{1}{2},
\end{equation}
determining the optimal measurement points $\tilde{\delta}_\pm$ to be given by
\begin{equation}
\tilde{\delta}_\pm = \pm \; \mathrm{arccos}(2^\frac{n-1}{n} - 1).
\end{equation}
Expanding $F(\delta)$ around $\varphi_\mathrm{opt} = 0$ up to first order further provides us with a compact linear expression to estimate $\varphi_\mathrm{opt}$
\begin{equation}
\varphi_\mathrm{opt} \simeq \frac{1}{\alpha} \; \frac{F(\tilde{\delta}_+) - F(\tilde{\delta}_-)}{F(\tilde{\delta}_+) + F(\tilde{\delta}_-)}
\end{equation}
with 
\begin{equation}
\alpha = n \;\mathrm{sin}(\tilde{\delta}_+) \;\bigg[\frac{1 + \mathrm{cos}(\tilde{\delta}_+)}{2} \bigg]^{n-1}.
\end{equation}
A graphical illustration of the 2-point phase estimation is shown in \fig{fig_fpe}(e).
After starting the optimization procedure with regular equally spaced phase stepping, once $\varphi_\mathrm{opt}$ starts to converge to small angles the 2-point measurement scheme can be employed.
In the measurements presented in Fig. 4 of the main text, the 2-point scheme was started after $N_\mathrm{mode}/2 = 128$ iterations, with the whole optimization running over $3N_\mathrm{mode} = 768$ iterations.
Using a 6-point estimation in the beginning, this strategy reduced the measurement time by factor 2.25 compared to keeping the initial scheme till the end.
Note that this technique is not limited to non-linear feedback and can also be applied in the linear case.
For an increased precision, also higher order terms of the expansion of $F(\delta)$ can be taken into account.

\end{document}